\documentstyle[11pt,moriond,epsfig]{article}

\def\li2{{\rm Li}_2}

\def\gev{{\rm GeV}}
\def\tev{{\rm TeV}}

\def\roughly#1{\,\,\raise.3ex\hbox{$#1$\kern-.75em\lower1ex\hbox{$\sim$}}\,\,}

\def\beq{\begin{equation}}
\def\eeq{\end{equation}}
\def\bea{\begin{eqnarray}}
\def\eea{\end{eqnarray}}

\def\prl#1#2#3{Phys. Rev. Lett. {\bf #1} (#2) #3}
\def\mpl#1#2#3{Mod. Phys. Lett. {\bf A#1} (#2) #3}

\def\gsim{{~\raise.15em\hbox{$>$}\kern-.85em
          \lower.35em\hbox{$\sim$}~}}
\def\lsim{{~\raise.15em\hbox{$<$}\kern-.85em
          \lower.35em\hbox{$\sim$}~}}

\def\Emiss{\not  \! \! E}
\def\jet{{\rm jet}}

\def\slashchar#1{\setbox0=\hbox{$#1$}           
   \dimen0=\wd0                                 
   \setbox1=\hbox{/} \dimen1=\wd1               
   \ifdim\dimen0>\dimen1                        
      \rlap{\hbox to \dimen0{\hfil/\hfil}}      
      #1                                        
   \else                                        
      \rlap{\hbox to \dimen1{\hfil$#1$\hfil}}   
      /                                         
   \fi}                                         %
\catcode`@=11
\long\def\@caption#1[#2]#3{\par\addcontentsline{\csname
  ext@#1\endcsname}{#1}{\protect\numberline{\csname
  the#1\endcsname}{\ignorespaces #2}}\begingroup
    \small
    \@parboxrestore
    \@makecaption{\csname fnum@#1\endcsname}{\ignorespaces #3}\par
  \endgroup}
\catcode`@=12

\def\jfig#1#2#3{
 \begin{figure}
\begin{centering}
 \epsfysize=2.8in
 \centerline{\epsfbox{#2}}
 \caption{#3}
 \label{#1}
\end{centering}
 \end{figure}}

\def\sch{Schwarzschild }
\def\eq#1{eq.~(\ref{#1})}
\def\beqa{\begin{eqnarray}}
\def\eeqa{\end{eqnarray}}

\def\be{\begin{equation}}
\def\ee{\end{equation}}
\def\bea{\begin{eqnarray}}
\def\eea{\end{eqnarray}}

\begin{document}
\begin{flushright}
CERN-TH/2002-114\\
hep-ph/0205265
\end{flushright}
\vspace*{4cm}
\title{ \large \bf Transplanckian collisions  at future accelerators 
\footnote{Talk given at the XXXVIIth Rencontres 
de Moriond,  Electroweak Interactions and Unified Theories, Les Arcs, March 9-16, 2002.}
}

\author{Riccardo Rattazzi }

\address{Theory Division, CERN\\ CH-1211, Geneva 23, Switzerland
}

\maketitle

\abstracts{Scattering at transplanckian energies offers model independent tests
of TeV scale gravity. Black-hole production is one spectacular signal, though a full
calculation of the cross section is not yet available. Another signal 
is given by gravitational elastic scattering, which is maybe less spectacular
but which can be nicely computed in the forward region using the eikonal approximation.
In this talk I discuss the distinctive signatures of eikonalized scattering at future accelerators.
}
 
The hierarchy $G_N/G_F\sim 10^{-33}$ between the Newton and Fermi constants
is a fact of Nature. This fact is considered by particle theorists to be a
 problem. This is because in the Standard Model (SM)
the Fermi scale $G_F^{-1/2}\sim 300$ GeV is determined by the vacuum expectation
value (VEV) of a scalar field, which is in turn determined
by its mass parameter. Scalar masses are known to be very sensitive to 
ultraviolet (UV) quantum corrections. In a theory where the fundamental
UV scale is $M_P\sim G_N^{-1/2}\sim 10^{19}$ GeV one would also expect 
a scalar mass of the same order of magnitude and consequently no hierarchy
between $G_N$ and $G_F$.

Until a few years ago all the efforts to explain the ratio 
$G_N/G_F\sim 10^{-33}$  focussed on the 
denominator: trying to explain why the Higgs mass is as small
as it is. In technicolor models the Higgs is naturally so light
because it is a composite particle at energies above the weak scale.
In supersymmetric models, on the other hand, the Higgs can be
elementary up to the Planck scale, as  the boson-fermion symmetry
protects its mass. Arkani-Hamed, Dimopoulos and Dvali (ADD)\cite{add} 
have instead suggested that one could formulate (and maybe solve) the problem
by focussing on the numerator $G_N$. They have proposed a scenario where the 
fundamental quantum gravity scale is of the order of the Fermi scale.
In  order to account for the observed weakness of gravity they assume
that there exists a number $n$ of new compact spacelike dimensions.
The relation between the microscopic Newton constant $G_D$, valid in the
4+n dimensional theory, and the macroscopic $G_N$, describing gravity
at distances larger than the compactification radius $R$, is
\beq
G_N=\frac{G_D}{R^n}.
\label{micromacro}
\eeq
Then for $G_D\sim (1 \,\tev)^{-2-n}$, the right value of $G_N$ is reproduced
for rather large compactification radii ranging from $R\sim 1$ fm for $n=6$ 
to $R\sim 1$ mm for $n=2$. Such large values of $R$ are not in contradiction
with present gravity experiments, as Newton's law as been tested only down to distances
just below a mm. On the other hand, gravity excluded, all the observed particles and interactions 
 are very well described by a 3+1 dimensional quantum field theory,
the Standard Model, down to length scales smaller than the Z boson Compton wavelength
$\lambda_Z\sim 10^{-3}$ fm. In order to account for this fact,
ADD assume that the SM degrees of freedom are localized on a  defect extending over
the 3 ordinary non-compact directions in space,
a 3-brane. The possibility to localize particles on defects, or submanifolds, was
remarked a while back in field theory \cite{Rubakov:bb} and is also naturally realized in 
string theory by Dirichlet-branes \cite{Polchinski:1996na}. A possible string realization of the ADD
proposal was first given in ref. \cite{aadd}.
Therefore as long as the size of the brane is somewhat smaller than the weak
scale, SM particles  behave as ordinary 3+1 dimensional degrees of freedom up to
the energies explored so far.

As it stands, the ADD proposal is not yet a solution of the hierarchy
problem, but a ``new guise of the problem''\cite{ahdmr}. Instead of the small
Higgs VEV of the old guise, we now need to explain the very large
value of the compactification volume
\beq
V_nM_D^{n}=R^nM_D^n\sim 10 ^{33}
\eeq
where $M_D^{2+n}=1/G_D$ is the fundamental gravity scale. As we are dealing with gravity, $R$
is a dynamical degree of freedom, a scalar from the point of view of 4 dimensions.
Since we want $\langle R\rangle$ much bigger than its natural scale $1/M_D$, the 
scalar potential $V(R)$ will have to be much flatter than naively expected
 at large values of $R$. As far as we understand, the most natural way to achieve
such flat potentials is by invoking supersymmetry. So, if the ADD scenario
is realized in Nature it is likely to be so together with supersymmetry at some stage.
Notice that in the conventional guise of the hierarchy problem supersymmetry is
invoked to ensure a flat potential (small mass) at small values of the Higgs field. 
Indeed we have mapped a small VEV problem into an essentially equivalent large
VEV problem. But notice that in the new scenario the hierarchy problem has become a sort of
cosmological constant problem. Indeed a vacuum energy density
$\Lambda^{4+n}$ would add to the radius potential a term $\sim \Lambda^{4+n}R^n$.
This grows very fast at large $R$ so we expect \cite{ahdmr} that $\Lambda^{4+n}$ should be much smaller than
its natural value $(\tev)^{4+n}$.
At this point one may caustically remark that we succeded in associating a problem
for which we had some solutions (technicolor, supersymmetry) to a problem for
which we have no convincing one. But this would probably be unfair, since bulk supersymmetry  does indeed
help in explaining the radius hierarchy, while it is seems unfortunately useless
to explain the smallness of the 4d cosmological constant.
Moreover, one may optimistically remark
that in the ADD scenario experiments, rather than theory, will shed light on all
the mistery of quantum gravity, including possibly the cosmological constant problem(s).
This is undoubtedly the reason why the ADD proposal is considered so interesting
by particle physicists.

There are two classes of laboratory tests of this scenario. The first is
given by the search for deviations from Newton's law at short but macroscopic distances. 
This is done in table top experiments. These deviations could be determined by
the light moduli, like the radius $R$ \cite{ahdmr}, or by the lowest Kaluza-Klein (KK) J=2
modes. At present no deviation as been seen down to a length $\sim 200\,\mu{\rm m}$ 
\cite{Adelberger:2002ic}. The second class of tests is given by high energy collisions \cite{everybody}. 
In this case we deal with either gravitons at virtuality $Q\gg 1/R$ or with real gravitons
measured with too poor an energy resolution to distinguish individual KK
levels (around mass $m$ the level separation is $\Delta m\sim 1/(R^n m^{n-1}$).
Therefore we can take the limit $R\to \infty$ and work as if our brane were embedded
in ($4+n$)-dimensional Minkowski space. Gravity couples with a strength $(E/M_D)^{n+2}$,
so we can distinguish three energy regimes. At $E\ll M_D$ we are in the cisplanckian regime.
Here gravity is weak, and the signals involve the emission of a few graviton quanta, which escape
 undetected into the  extra dimensions.
 Interesting examples
are $e^+e^-\to \gamma+{\rm graviton}=\gamma+\Emiss$ or $pp \to \jet +\Emiss$ or
 even the  invisible decay of the Higgs
into just one graviton \cite{grw2}. Processes with real graviton emission can be predicted
in a model independent way  in terms of just one parameter $M_D$ (only in the case of Higgs decay
another parameter enters). On the other hand virtual graviton exchange
is dominated already at tree level by uncalculable UV effects. These amplitudes are therefore
associated to a new class of contact interactions whose coefficients depend on the details
of the fundamental theory of quantum gravity.
 The second regime is the planckian one where $E\sim M_D$, which would give
experimental access to the theory of quantum gravity. The signals are here highly model dependent,
meaning that this is the regime where the most relevant experimental information will be extracted.
If string theory is at the core, then one characteristic signal is given by the observation
of Regge excitations \cite{peskin}. Finally there is the transplanckian regime $E\gg M_D$,
which is the subject of this talk. Though very naively one would expect things
to be completely out of control, in the transplanckian regime gravity becomes rather simple:
it is basically described by classical general relativity. As such, the transplanckian
regime, like the cisplanckian, can offer model independent tests of the large extra dimension scenario.
 
To better understand the transplanckian regime it is useful to do some dimensional analysis working
in units where $c=1$ but keeping $\hbar\not = 1$. Using 
$G_D$ (($[G_D]={\rm length}^{n+1}E^{-1}$) and the center of mass (c.m.) energy $\sqrt s$ we have
\beq
M_D^{n+2}=\hbar^{n+1}/G_D,\quad\quad\quad\lambda_P^{n+2}=\hbar G_D,\quad\quad\quad\lambda_B=\hbar/{\sqrt s}
\label{dimensional}
\eeq
where the  Planck length $\lambda_P$ represents the length below which quantum gravity
fluctuations of the geometry are important, while $\lambda_B$ is the de Broglie wavelength
of the scattering quanta in the c.m.. 
Combining $G_D$ and $\sqrt{s}$, we can form the \sch radius of a
system with c.m. energy $\sqrt{s}$~\cite{perry}
\beq
R_S =\frac{1}{\sqrt{\pi}}
\left[ \frac{8\Gamma \left( \frac{n+3}{2}\right)}{(n+2)}
\right]^{\frac{1}{n+1}}~ \left( {G_D \sqrt{s}}\right)^{\frac{1}{n+1}}.
\label{rschwarzschild}
\eeq
This is the length at which curvature effects become significant.
In the limit $\hbar \to 0$, with $G_D$ and $\sqrt s$ fixed,
$M_D$ vanishes,
showing that classical physics correspond to transplanckian 
(macroscopically large) energies. Moreover, in the same limit,
$R_S$ remains finite, while
the two length scales $\lambda_P$ and $\lambda_B$ go to zero.
Therefore, the transplanckian regime corresponds to a classical 
limit in which
the length scale $R_S$ characterizes the dynamics,
\beq
\sqrt{s} \gg M_D  ~~~~~~~\Rightarrow~~~~~~R_S\gg \lambda_P \gg \lambda_B.
\eeq
We can see this more explicitly by considering the classical
scattering angle for a collision with impact parameter 
$b$. By simple dimensional arguments it is
$\theta\sim G_D\sqrt s/b^{n+1}=(R_S/b)^{n+1}$.
 This shows
that by increasing $\sqrt s$ we can obtain a finite
$\theta$ by going to large $b$, where
short distance quantum gravity effects are suppressed. More precisely,
in order to describe the collision classically, two conditions must
be met: {\it i}) $b\gg \lambda_P$ in order to suppress quantum gravity fluctuations;
{\it ii}) $\theta L/\hbar=\theta b{\sqrt s}/\hbar\gg 1$
\cite{landau} in order to suppress ordinary quantum mechanical effects
due to the ondulatory nature of the colliding particles. This second requirement
corresponds to $b^n \ll G_Ds/\hbar\equiv b_c^n$. In \eq{dimensional} we have knowingly disregarded
$b_c$ as it is related to ordinary quantum mechanical effects. It corresponds to the
critical impact parameter above which the classical scattering angle becomes
smaller than its quantum indetermination. (The presence of a $b_c$ in potential scattering is
a known property of potentials vanishing faster than $1/r$ at infinity: notice indeed that in
our case $b_c$ is only defined for $n>0$ corresponding to a $1/r^{1+n}$ potential.)
Now, for $\sqrt s \gg M_D$ we have $\lambda_P\ll R_S \ll b_c$, so that there is a range of 
impact parameters where the motion is well described by classical physics.

This property of gravity should be contrasted to what happens in the case of 
gauge interactions mediated by vector bosons. In gravity the role
of charge is played by energy, so with just one incoming quantum we can have a macroscopic 
source of gravity if $E=\hbar \nu\gg M_D$. In the case of gauge interactions the charge
of one fundamental quantum is $\hbar$ times a number which we can conventionally set to 1
by rescaling the gauge field. Then in order to have a macroscopic source we need an
object like a nucleus or a soliton involving many charged quanta. Consider the angle
for the scattering between two objects carrying $Z$ units of charge 
$\theta=g^2\hbar^2 Z^2/{\sqrt s}b^{n+1}$, 
where the gauge coupling has dimension $[g^2]={\rm length}^{n-1}E^{-1}$ and
$Q=\hbar Z$ is the charge. The conditions for classical motion are  as before.
{\it i}) is replaced by $b\gg \lambda_g=(\hbar g^2)^{1/n}$: $\lambda_g$ is the typical length scale
where a gauge theory in 4+n dimensions becomes strongly coupled. Simoultaneous satisfaction of
{\it i}) and and {\it ii}) implies $Z\gg 1$, which excludes the scattering
of elementary quanta. 

The physics of transplanckian collisions was studied in a series of papers more than ten
years ago \cite{thooft,amati,muzinich,verlinde}. String theory corrections to the classical
result were even considered \cite{amati}. The basic picture is that for impact parameter
$b\gg R_S$, the particles scatter by a small angle $\theta \sim (R_S/b)^{n+1}$ while for even
larger $b>b_c$ the classical angle is so small that ordinary quantum mechanical effects come into play.
In the $b\gg R_S$ regime non linear effects due to the superposition of the gravitational fields
of the two scatterers are small so one can work with linearized gravity. Moreover since the scattering
is at small angle the amplitude can be calculated by using the eikonal approximation.
The eikonal amplitude can be  obtained in two equivalent ways\cite{kab1}. In one approach what is 
calculated is the phase shift of
the wave function of one particle when crossing the gravitational shockwave field created by the other
particle \cite{thooft}. The other approach \cite{amati,muzinich} consists in the direct resummation of the 
series of graviton exchange ladder 
diagrams \cite{eik}. The eikonal approximation breaks down for impact parameters $b\sim R_S$, where the
scattering angle becomes $O(1)$. No full calculation in this regime is available right now.
A reasonable expectation is that for $b\lsim R_S$ the two particles, with most of their energy,
collapse to form a black hole (BH). Heuristically, this is because at the moment the particles cross 
there is an amount of energy $\sqrt s$ localized within a radius $b< R_S$ so that gravitational
collapse should follow. More rigorously, following original unpublished work by Penrose on head-on
 collisions ($b=0$),
it has been recently proven \cite{Eardley:2002re} that for small enough impact parameter a marginally 
trapped surface
forms at the overlap between the two gravitational shockwaves. Then by the
singularity theorems \cite{hawkingellis} a horizon should form. The study of 4-dimensional
$b=0$ collisions \cite{d'eathpayne} shows that only a small fraction ($\sim 20 \%$) of the original energy
is radiated away in gravitational waves. It is reasonable to assume that the 4+n-dimensional case does 
not differ significatively. Based on this assumption the cross section for black-hole production is estimated on
simple geometrical grounds to be $\sigma_{BH}\sim\pi R_S^2$ where $R_S$ is given in \eq{rschwarzschild}.
Pending a full calculation, the phenomenological studies so far have just taken $\sigma_{BH}=\pi R_S^2$,
expecting that the correct result will not be much different (see ref. \cite{voloshin} for some criticism).

If the large extra dimension scenario is realized in nature with $M_D\sim 1$ TeV, then 
(maybe optimistically) LHC with its 14 TeV c.m. energy may start probing physics in the transplanckian 
regime\cite{banks,blackh,empa,ratta,us}.
Of course a machine with $O(100)$ TeV c.m. energy like VHLC \cite{vlhc} would require less optimism.
In the remaining part of the talk I will outline what are the signatures of gravitational elastic
scattering and also, briefly, black-hole production at the LHC. The results can easily be
generalized to higher energy machines.

Consider elastic scattering first. Since the dynamical regime we are focusing on overlaps with the classical
limit where the action $S/\hbar\gg 1$, the amplitude is not given by a perturbative calculation.
In other words, the classical limit implies exchange or emission of a large (infinite) number 
of graviton quanta, so we cannot do with a finite number of Feynman diagrams. In the forward region, however,
this infinite set is consistently given by the series of ladder and crossed ladder diagrams. The result
does not depend on the spin of the particles, since in the eikonal limit the particle line is taken on
shell and the coupling to the graviton is simply given by $\langle T_{\mu\nu}\rangle=p_\mu p_\nu$,
where $p$ is the incoming 4-momentum. In the forward region
 the momentum transfer $q$ is basically given by the two-dimensional tranverse momentum $q_\perp$: 
$t =q^2\simeq -q_\perp^2$. The series of ladder diagrams nicely exponentiates. The resulting amplitude
is more conveniently written by trading $q_\perp$ for its Fourier conjugate variable,
the impact parameter 2-vector $b$,
\beq
{\cal A}_{\rm eik}={\cal A}_{\rm Born}(q_\perp^2)+{\cal A}_{\rm 1-loop} (q_\perp^2)+\dots
=-2is \int d^2b e^{iq_\perp b} \left( e^{i\chi}
-1\right) 
\label{eicon}
\eeq

\beq
\chi (b_\perp )=\frac{1}{2s} \int \frac{d^2q_\perp}{(2\pi)^2}
e^{-iq_\perp b}{\cal A}_{\rm Born}(q_\perp^2).
\label{eikp}
\eeq
$\chi(b)$ is called the eikonal phase and
$e^{i\chi}$ represents the amplitude in impact parameter space. Unitarity at
small angle is thus manifestly satisfied.
Notice that we work with a two-dimensional transferred momentum since the scattered particles
live on a 3-brane. On the other hand the exchanged gravitons are $D$-dimensional. So the Born amplitude
involves a sum over the n-momentum $k_\perp$ tranverse to the brane \cite{everybody}. This  sum is known 
to be UV divergent
for $n\geq 2$. These divergences can be parameterized at low energy by a set of contact interactions.
Naively, they would give $\delta$-function contributions localized at $b=0$ to $\chi(b)$.  
 On physical grounds, however we expect these local divergences to be softened
by the fundamental theory of gravity at some finite but small,
impact parameter $b\sim \lambda_P$
(or, more likely, at $b$ of the order of the string length $ \lambda_s$). 
These short-distance effects should plausibly give rise to 
${\cal O}(\lambda_P^2)$
corrections to the cross section, while, as we will see shortly, the 
long-distance eikonal amplitude gives a cross section 
that grows with a power of $\sqrt s$,
thus dominating at large energies. Then we only need to focus on the non-local calculable piece in $A_{\rm Born}$.
For this purpose it is convenient to use dimensional regularization
\beq
{\cal A}_{\rm Born} (-t)=\frac{s^2}{M_D^{n+2}} \int \frac{d^n k_T}{t-k_T^2}
=\pi^{\frac{n}{2}} \Gamma(1-n/2)\left( \frac{-t}{M_D^2}\right)^{\frac{n}{2}
-1}\left( \frac{s}{M_D^2}\right)^2,
\label{born}
\eeq
from which we get the eikonal phase

\beq
\chi = \left( \frac{b_c}{b}\right)^n,
\quad\quad\quad
b_c\equiv \left[ \frac{(4\pi)^{\frac{n}{2}-1}s\Gamma (n/2)}{2M_D^{n+2}}
\right]^{1/n}.
\label{bbc}
\eeq
Notice that by inserting this result for $\chi$ in the integral in
\eq{eicon} we obtain an ultraviolet 
finite result. This is so, even though the contributions to \eq{eicon}
from the individual terms
in the expansion $e^{i\chi}=1+i\chi+\dots$ are ultraviolet 
divergent, corresponding
to the fact that each individual Feynman diagram in the ladder expansion is 
ultraviolet
divergent but the complete sum is finite. 
Moreover, since $\chi\propto b^{-n}$, the integrand in \eq{eicon}
oscillates very rapidly as $b\to 0$, showing that the ultraviolet 
region gives but a small contribution to the amplitude.
Replacing \eq{bbc} into \eq{eicon},  
the momentum space amplitude~\cite{ratta} is written in terms of Meijer's G-functions.  

The length scale $b_c$, as expected from the general discussion at the beginning,
controls ordinary quantum mechanical effects. For $b\ll b_c$ the eikonal phase
is large and rapidly oscillating, corresponding to the classical limit. For $b\gsim b_c$ the phase
is small and quantum mechanics sets in. These two regimes are realized in momentum space as follows.
For semi-hard momenta $\sqrt{s}\gg q \gg b_c^{-1}$
the integral in \eq{eicon} is dominated by the
stationary-phase value of the impact parameter $b_s \equiv b_c (n/qb_c)^{\frac{1}{n+1}}<b_c$. 
This is precisely the classical region. Here the concept of trajectory makes sense and the scattering angle
is
\beq
\theta_{\rm cl}=-\frac{\partial \chi}{\partial L}=\frac{2n\Gamma (n/2)}{
\pi^{n/2}}~\frac{G_D\sqrt{s}}{b^{n+1}}.
\label{einsa}
\eeq
In the limit $n\to 0$, we recover the Einstein angle $\theta_{\rm cl}
=4G_D\sqrt{s}/b$
 while, for $n>0$, \eq{einsa} gives its higher-dimensional 
generalization (for the case in which also the sun moves at ultrarelativistic speed in the c.m.!)

In the soft region  $ q \lsim b_c^{-1}$,
the integral in \eq{eicon} is dominated by $b$ of the
order of (or slightly smaller than) $b_c$. This means that the eikonal
phase $\chi=(b_c/b)^n$ is of order one and the quantum
nature of the scattering particles is important (although quantum gravity effects are 
negligible and the exchanged graviton is treated as a classical field).
Moreover, notice that the relevant $\chi$ never becomes much smaller than 1,
and therefore we never enter the perturbative regime in which a loop
expansion for the amplitude applies. Even though the interaction vanishes
at $b\to \infty$ (where $\chi \to 0$), we never reach the Born limit.
Even for $q=0$, the scattering is dominated by $b\sim b_c$ and not 
by $b= \infty$, as opposed to the Coulomb case. 
This result follows from the different dimensionalities of the spaces on which
the scattered particles and exchanged graviton live. It does not hold
for the scattering of bulk particles. In that case the eikonal phase is
unchanged, but the impact parameter vector $b$ becomes $(2+n)$-dimensional. In particular
$d^2b\to d^{2+n}b$ in   \eq{eicon} so that for $q=0$, the integral is infrared
dominated by large values of $b$ and
quadratically divergent (for any $n$). Indeed one finds ${\cal A}_{\rm eik}
(q\to 0) =A_{\rm Born}\sim b_c^n /q^2$, encountering the Coulomb singularity
characteristic of long-range forces. 

At the LHC, the observable of interest is jet-jet production at small
angle (close to beam) with large center-of-mass collision energy \cite{us}.
The scattering amplitude is the same for any two partons.  
The total jet-jet 
cross-section is then obtained
by summing over all possible permutations of initial state quarks
and gluons, using
the appropriate parton distribution weights and enforcing kinematic
cuts applicable for the eikonal approximation.

Defining $\hat s$ and $\hat t$ as Mandelstam variables of the parton-parton
collision,  
we are interested
in  events that have $\sqrt{\hat s}/M_D\gg 1$ and 
$-\hat t/\hat s \ll 1$. We can extract $\sqrt{\hat s}$ 
from the jet-jet invariant mass $M_{jj}=\sqrt{\hat s}$, 
and $\hat t$ from the rapidity separation of
the two jets $-\hat t/\hat s=1/(1+e^{\Delta \eta})$, where 
$\Delta \eta \equiv \eta_1-\eta_2  = \ln(1+\cos\hat\theta)/(1-\cos\hat\theta)$
and $\theta$ is the c.m. scattering angle.
The kinematical region of interest is defined by
the equivalent statements 
\beq
\Delta\eta\to \infty ~~~\leftrightarrow ~~~ 
\hat\theta \to 0 ~~~ \leftrightarrow ~~~
\frac{-\hat t}{\hat s}\to 0.
\eeq
Since the partonic scattering probes a region of size $b$ inside the protons, it is reasonable
to evaluate the parton-distribution functions at the scale $Q^2=b_s^{-2}$ if $q>b_c^{-1}$
and $Q^2=q^2$ otherwise ($q^2\equiv -\hat t$) \cite{ratta}.
In the computations the CTEQ5~\cite{cteq} parton-distribution functions have been used.
The SM di-jet cross section has been computed using Pythia~\cite{pythia},
ignoring higher-order QCD corrections. For simplicity
the background is defined as the jet-jet cross-section from QCD with
gravity couplings turned off, and the signal as the jet-jet cross-section
from the eikonal gravity computation with QCD turned off. In reality, 
SM and gravity contributions would be simultaneosly present. However in the  
interesting kinematic region  gravity dominates, so this simple approach
is adequate.

The di-jet differential cross section $d\sigma_{jj}/d|\Delta\eta |$
is plotted in
fig.~\ref{fig-dely} for $n=6$, 
$M_{jj}>9\,\tev$ and $M_D=1.5\,\tev$ and $3\,\tev$. Similar plots for different $n$ can be found in ref. \cite{us}.
Since the parton-distribution functions 
decrease rapidly at higher $M_{jj}$, the plot is dominated by events with
$M_{jj}\sim 9\,\tev$. Notice the peak structure at intermediate values of $\Delta y$,
corresponding to impact parameters of order $b_c$ (evaluated at ${\sqrt s}= 9\,\tev$), 
{\it i.e.} to the transition region between classical and quantum 
mechanical scattering. These peaks represent the diffraction of waves scattered around $b\sim b_c$.
They are a characteristic feature of the higher-dimensional gravitational field which, while being of infinite range,
behaves somewhat like a potential well of size $b_c$. In the Coulomb case ($n=0$), such a length scale does not exist 
and therefore no diffractive pattern is produced.
\jfig{fig-dely}{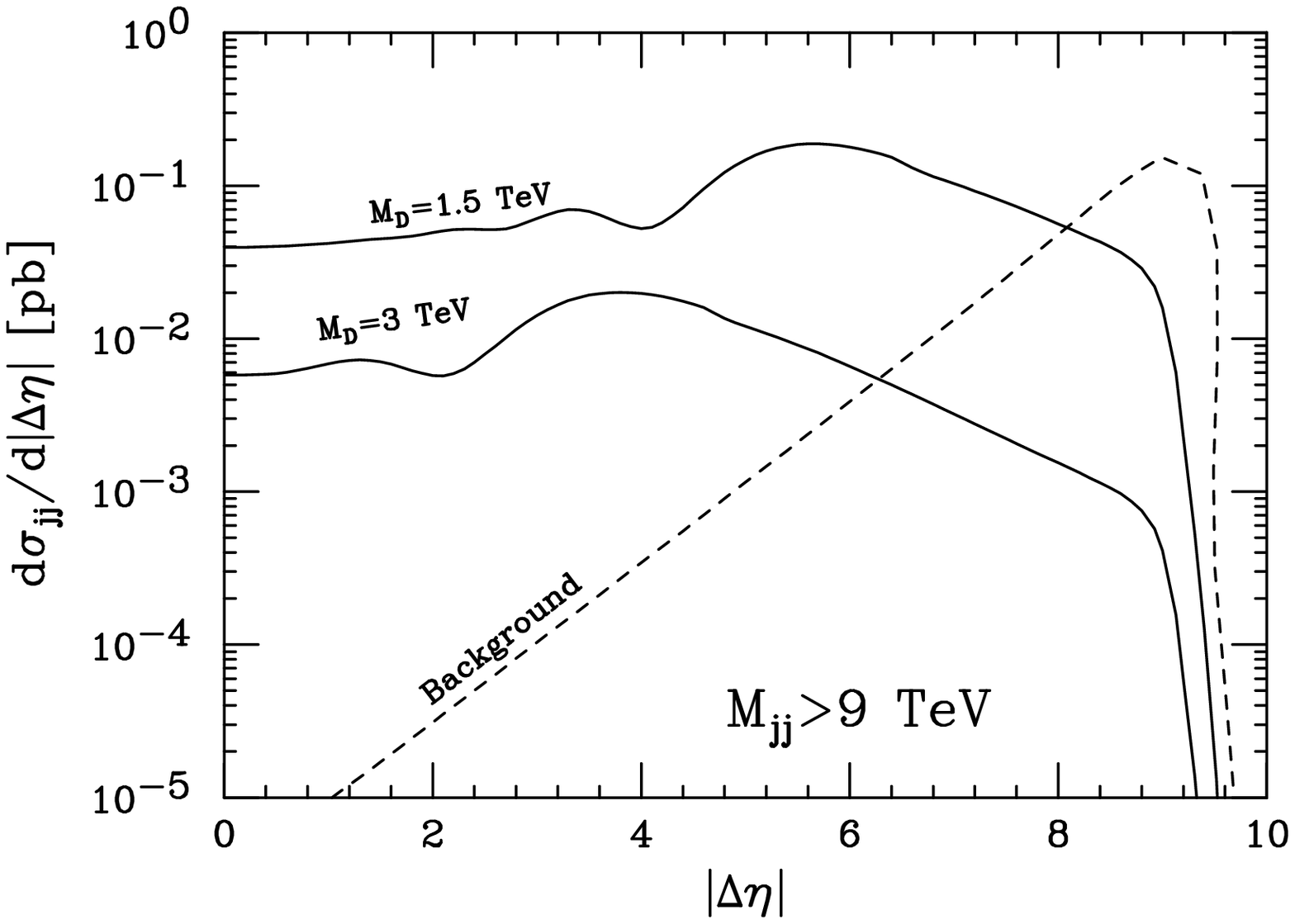}{The 
di-jet differential cross section 
$d\sigma_{jj}/d|\Delta\eta |$ from
eikonal gravity for $n=6$, 
$M_{jj}>9\tev$, when both jets have $|\eta | <5$ and $p_T>100\gev$,
 and for $M_D=1.5\tev$ and $3\tev$.  The
dashed line is the expected rate from QCD.}


 
Since the two jets are experimentally indistinguishable, I have used
$|\Delta \eta |$, instead of $\Delta \eta$, as the appropriate
kinematical variable to plot. This means that the experimental signal
considered here contains also contributions from scattering with large
and negative $\Delta \eta$, which corresponds to partons colliding
with large momentum transfer and retracing their path backwards.
For the background, these effects are calculable and taken into account.
However, the theoretical estimate of the signal at negative $\Delta\eta$
lies outside the range of validity of the eikonal approximation. Indeed the region of
$\theta \sim \pi$ corresponds to impact parameters $\sim R_S$. Its
contribution to the differential cross section will be $d\sigma/d t\sim \pi R_S^2/s$, {\it i.e.} 
parametrically smaller
than the forward one $d\sigma/d t\sim \pi b_c^4$. Therefore it is expected to be negligible and
can be safely ignored. 

As is evident from the figure the gravitational cross section is harder that
the QCD one. This is because the latter is dominated by the forward Coulomb singularity,
while the forward eikonal amplitude is finite. In the semi-hard region $q\gsim b_c^{-1}$ the gravitational
cross section is also much bigger than the QCD one: as we discussed before, large cross sections
at large energy and finite angle are a clear signal of
gravitational interactions. This is because energy itself plays the role of charge in gravity. 
It is difficult to imagine some other physics that mimics this result.

To get an idea of the sensitivity at LHC one can study the total integrated cross section
using some illustrative cuts.
To stay in the small angle region while beating the QCD background a reasonable choice is
$3<\Delta \eta<4$.
The integrated cross section as a function of minimum jet-jet invariant
mass is  shown in fig.~\ref{fig-Mjj}. 
This plot shows the important feature that
the signal cross-section is flatter in $M_{jj}$ than the background.
This enables better signal to background for larger $M_{jj}$ cuts,
which is the preferred direction to go for the 
validity of the transplankian limit.  Therefore, one should make 
the largest possible $M_{jj}$ cut that still
has a countable signal rate for a given luminosity. For an integrated luminosity of
$30 \,{\rm fb}^{-1}$, corresponding to expectations for one year of running, 
the plot of fig.~\ref{fig-Mjj} shows that several hundreds to thousands of events can
plausibly be expected.
\jfig{fig-Mjj}{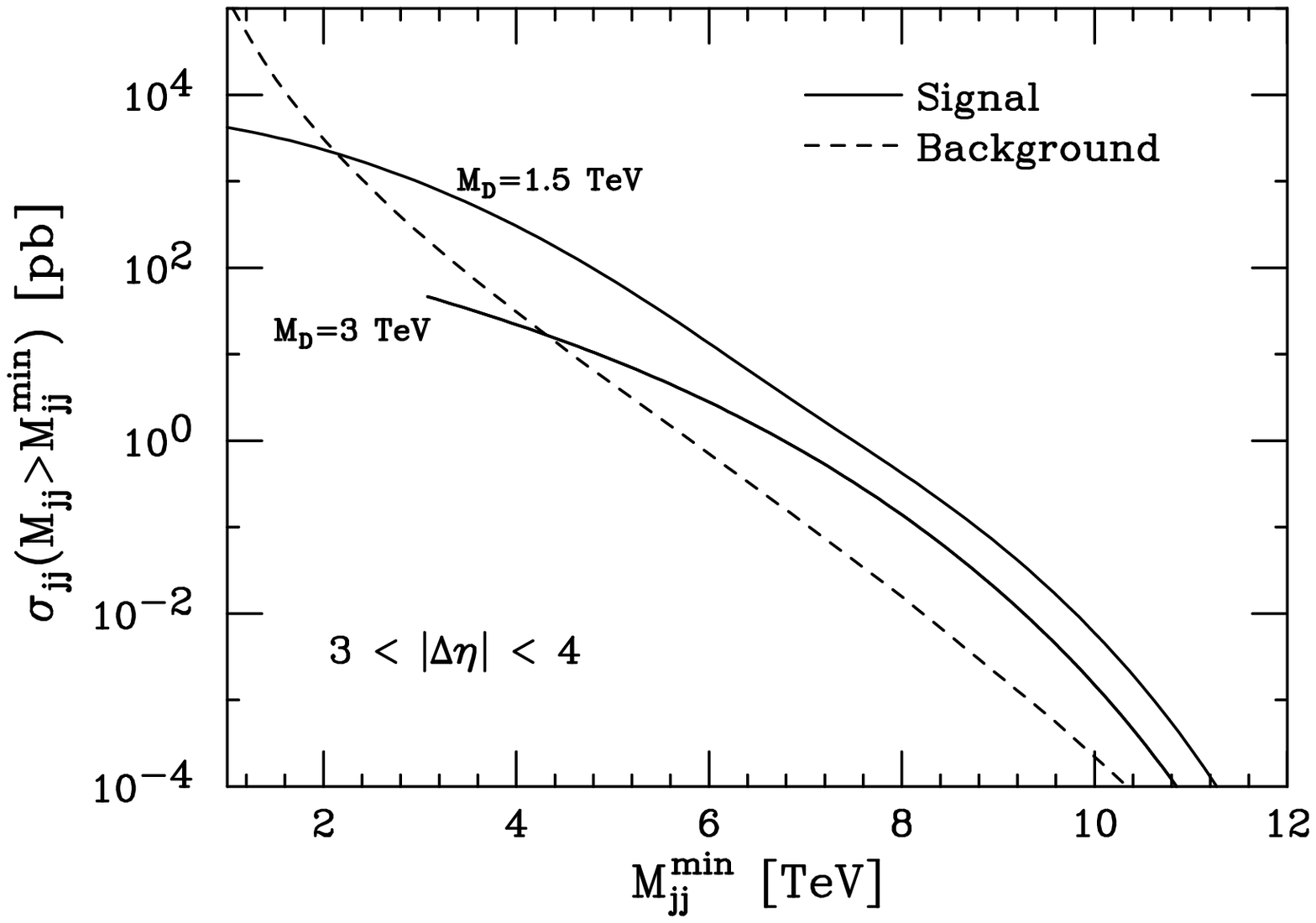}{Total 
integrated di-jet cross-section
for $3< |\Delta \eta| <4$, $n=6$, and $M_{jj}>M_{jj}^{\rm min}$,
when both jets have $|\eta | <5$ and $p_T>100\gev$. Lines are plotted
for $M_D=1.5$ and 3 TeV.
The eikonal approximation 
is reliable only where $M_{jj}/M_D\gg 1$. The expected QCD
rate is given by the dashed line.}

One can also study (see ref. \cite{us}) the sensitivity reach at LHC
considering the total integrated
cross-section as a function of $M_D$ for $M_{jj}>3M_D$ (optimistic)
and $M_{jj}>6M_D$ (more conservative). One finds that in the two cases the reach for $M_D$ is
respectively $3.5$ and $1.8\,\tev $ almost independent of $n$.   

Similar sensitivities are obtained for the more spectacular events with black-hole production \cite{blackh,us}.
As we said, here a full calculation is not avaliable, but the order of magnitude estimate
leads to a cross section for BH production at LHC which, for $M_D\lsim 3 \,\tev$,  can range  from
$10^{-2}$ to $10^2 {\rm pb}$. With $30 {\rm fb}^{-1}$ per year, LHC could then see several hundreds
(or even millions, if one is optimistic) of BH events. These objects happily decay very fast by
Hawking radiation with a temperature $T_H=(n+1)/(4\pi R_S)$. So one should not worry about their growing
by eating up the detector! A simple estimates shows that in order to produce such environmentally
dangerous BH's a c.m. energy in excess of $10^{19}\gev$ is needed. The black holes  decay with 
comparable probability
to any particle living either on the brane or in the bulk \cite{Emparan:2000rs}. 
If only gravity propagates in the bulk, then they will essentially decay on the brane
as it hosts a large  number ($\sim 100$) of degrees of freedom. The lifetime of a BH with mass
$M$ is $\tau \sim M_D^{-1}(M/M_D)^{(3+n)/(1+n)}$ while the multiplicity of the decay products
(mostly quarks and gluons giving rise to jets) scales like $(M/M_D)^{(2+n)/(1+n)}$.
So at the LHC the multiplicity could be of order $10\div 100$. As the parton distribution functions 
decrease rapidly at large 
$x$ BH's of a given mass are most often produced with a small boost. So the characteristic
of BH events is high multiplicity with high spherocity \cite{blackh}. A signal which has practically no background.

Elastic scattering and BH production may also affect the physics of cosmic rays \cite{Feng:2001ib,ratta} 
(see also ref. \cite{nussi}).
This is because they can lead to a significative enhancement
of the cross section of cosmic ultra high energy neutrini with nucleons in the atmosphere.
It is known that there should exist a cosmogenic neutrino flux, originated by
the inelastic scattering of primary protons on the cosmic microwave background. 
The c.m. energy of the neutrino nucleon system is $\sim \sqrt{(E_\nu/10^6\gev)}\,\tev$,
which could well be in the transplanckian regime for the ultra-energetic neutrini
with $E_\nu\sim 10^{10}\gev$. The Standard Model cross section dominated by W-boson exchange
is roughly given by \cite{Gandhi:1998ri} $\sigma_{SM}=10^{-5}(E_\nu/10^{10}\,\gev)^{0.363}$ mb.  The production 
of BH and elastic gravitational
scattering could lead to an enchancement of about $10^2$ of the cross section. As a result one expects
a similar enhancement in the rate of deeply penetrating horizontal showers. Future detectors
with improved sensitivity may be able to detect such events \cite{Ringwald:2001vk,Anchordoqui:2001cg}. 
The eikonalized neutrino nucleon cross section at small angle could even become of the order of $1$ mb.
Such a large cross section starts being interesting if one wants to explain the vertical
ultra GZK events as due to cosmic neutrini \cite{Weiler:2000ku}. However the eikonal cross section is soft,
corresponding to a very small energy transfer to the shower, so it cannot explain 
the ultra GZK events \cite{ratta}.

The above discussion neglects quantum gravity effects.  On general grounds and by analyticity in 
the transferred momentum  we expect
the corrections to elastic scattering in the interesting region $b\sim b_c$ to be of order
$(\lambda_P/b_c)^2$. Then by selecting events with $M_{jj}>6 M_D$ we expect $O(5\%)$
effects which is fairly good, while for $M_{jj}>3 M_D$ the effect can go up to $20\%$. However
it is possible that at LHC the effects are bigger. First of all, transplanckianity
at LHC requires $M_D\lsim 2-3\, \tev$. Such a low value is consistent with present direct
bounds on $M_D$ from direct graviton production. On the other hand one may in general expect  other 
quantum gravity effects, and in particular contact 4-fermion interactions of dimension 6. The presence of
these operators, given the bounds from LEP2, can push the lower bound \cite{benakli} on $M_D$ to above $\sim 4\, \tev$,
in which case LHC could not be considered a transplanckian machine. So our LHC study
 is truly based on the assumption that such dimension 6 terms are mildly suppressed. Indeed
there are examples in this direction in string realizations of the braneworld \cite{peskin}.
A second remark precisely concerns the case in which string theory is the theory of quantum gravity.
Then $\lambda_s=1/M_s$ and not $\lambda_P$ controls the  onset
of quantum gravity effects. In perturbative string theory (for instance \cite{aadd} type I) we have 
$(\lambda_P/\lambda_s)^{2+n}=\pi g_s^2<1$. Here $g_s$ is the string coupling which
is related to the gauge coupling by $g_s\sim 2\alpha$. We can assume such a relation to work qualitatively
in the realistic case with $\alpha$ taken to be some average between $\alpha_2$ and $\alpha_3$. Then we expect
a separation $\lambda_s>\lambda_P$ which means less separation between $b_c$ and $\lambda_s$ and bigger
quantum gravity effects. The previous naive estimate
of quantum gravity effects is enhanced by a factor $(1/4\pi \alpha^2)^{2/n}$. Taking $\alpha\sim 0.1$ we find
that for $n=2$ string effects could be $100\%$ while for $n=6$ they may conceivably be less than $20\%$.
Of course we do not want to take these estimates too seriously, as we truly need a full braneworld model
to calculate them. Else we should wait LHC and see first of all if there is a signal and then decide
how well it is explained by transplanckian scattering. Physically we expect string effects to suppress
scattering at large angles \cite{amati}, {\it i.e.} angles $\theta$ corresponding to impact parameters 
$b\lsim \lambda_s$. So in the plot of fig. \ref{fig-dely} the cross section at lower rapidity would be depleted.
Anyway, even if quantum effects at LHC will be large, it is important to have clear what the features of 
the asymptotic transplanckian regime are as they provide a benchmark with which to compare the data.
Notice, in this respect that elastic scattering is better off than black-hole production, since $b_c\propto
s^{1/n}$ is parametrically bigger (grows faster with $s$) than $R_s\sim s^{1/(2n+2)}$. 
If TeV gravity is truly  realized in nature then VLHC~\cite{vlhc},
whose center-of-mass energy for proton-proton collisions is envisaged
between  50 TeV and 200 TeV, would probably be a better  place to study transplanckian effects.
For instance, by assuming $\sqrt s=100\, \tev$, $M_s=3\, \tev$ and choosing an ``average'' value 
$\alpha = 0.05$ one finds that string effects parameterized by $\lambda_s^2/b_c^2$ are of order $5\%$. On
the other hand, for the same choice of parameters, one finds that the parameter $\lambda_s^2/R_S^2$ controlling
string corrections to black-hole production is still of order 1 for $n=2$ and $\sim 20\%$ for $n=6$.
So it is possible that even at VLHC the production of black holes is more appropriately replaced
by the production of string balls \cite{dimemp}.

In summary, transplanckian scattering offers a model independent test of theories with a low
gravity scale. The main processes are black-hole production and elastic scattering. The elastic cross section
is parametrically larger than the one for black holes and moreover
can be nicely calculated in the forward region. At the moment the black-hole
production cross section can only be estimated by dimensional analysis. The observation of a cross section
at finite angle growing with a power of $s$ would be a clean signal that 
the high-energy dynamics of gravity has been detected. If we are  lucky and $M_D$ is low enough, then
these signals may already show up at the LHC. Otherwise the discovery modes at LHC are graviton
emission, showing up as jet plus missing energy (roughly \cite{everybody} for $4\,\tev <M_D<8\,\tev$),
or the production of Regge excitations \cite{peskin}. Anyway if low scale gravity is discovered
at LHC, transplanckian scattering will very likely be studied at the VLHC.

I would like to thank R. Emparan, G.F. Giudice, M. Masip and J.D. Wells
for their crucial collaboration, and
J. March-Russell, P. Tinyakov and G. Veneziano for very useful conversations. 
I would also like to thank the organizers of this meeting
for the warm hospitality and the stimulating environment they provided.

\def\ijmp#1#2#3{{\it Int. Jour. Mod. Phys. }{\bf #1~}(19#2)~#3}
\def\pl#1#2#3{{\it Phys. Lett. }{\bf B#1~}(19#2)~#3}
\def\zp#1#2#3{{\it Z. Phys. }{\bf C#1~}(19#2)~#3}
\def\prl#1#2#3{{\it Phys. Rev. Lett. }{\bf #1~}(19#2)~#3}
\def\rmp#1#2#3{{\it Rev. Mod. Phys. }{\bf #1~}(19#2)~#3}
\def\prep#1#2#3{{\it Phys. Rep. }{\bf #1~}(19#2)~#3}
\def\pr#1#2#3{{\it Phys. Rev. }{\bf D#1~}(19#2)~#3}
\def\np#1#2#3{{\it Nucl. Phys. }{\bf B#1~}(19#2)~#3}
\def\mpl#1#2#3{{\it Mod. Phys. Lett. }{\bf #1~}(19#2)~#3}
\def\arnps#1#2#3{{\it Annu. Rev. Nucl. Part. Sci. }{\bf #1~}(19#2)~#3}
\def\sjnp#1#2#3{{\it Sov. J. Nucl. Phys. }{\bf #1~}(19#2)~#3}
\def\jetp#1#2#3{{\it JETP Lett. }{\bf #1~}(19#2)~#3}
\def\app#1#2#3{{\it Acta Phys. Polon. }{\bf #1~}(19#2)~#3}
\def\rnc#1#2#3{{\it Riv. Nuovo Cim. }{\bf #1~}(19#2)~#3}
\def\ap#1#2#3{{\it Ann. Phys. }{\bf #1~}(19#2)~#3}
\def\ptp#1#2#3{{\it Prog. Theor. Phys. }{\bf #1~}(19#2)~#3}


\begin{thebibliography}{99}


\bibitem{add} N.~Arkani-Hamed, S.~Dimopoulos and G.~R.~Dvali,
Phys.\ Lett.\ B {\bf 429}, 263 (1998).

\bibitem{Rubakov:bb}
V.~A.~Rubakov and M.~E.~Shaposhnikov,
Phys.\ Lett.\ B {\bf 125} (1983) 136.

\bibitem{Polchinski:1996na}
J.~Polchinski,
arXiv:hep-th/9611050.

\bibitem{aadd} I.~Antoniadis, N.~Arkani-Hamed, S.~Dimopoulos and G.~R.~Dvali,
Phys.\ Lett.\ B {\bf 436}, 257 (1998).

\bibitem{ahdmr}
N.~Arkani-Hamed, S.~Dimopoulos and J.~March-Russell,
Phys.\ Rev.\ D {\bf 63}, 064020 (2001)


\bibitem{Adelberger:2002ic}
E.~G.~Adelberger  [EOT-WASH Group Collaboration],
arXiv:hep-ex/0202008.

\bibitem{everybody} G.~F.~Giudice, R.~Rattazzi and J.~D.~Wells,
Nucl.\ Phys.\ B {\bf 544}, 3 (1999); E.~A.~Mirabelli, M.~Perelstein and M.~E.~Peskin,
Phys.\ Rev.\ Lett.\  {\bf 82}, 2236 (1999);T.~Han, J.~D.~Lykken and R.~J.~Zhang,
Phys.\ Rev.\ D {\bf 59}, 105006 (1999);
J.~L.~Hewett, 
Phys.\ Rev.\ Lett.\  {\bf 82}, 4765 (1999).

\bibitem{grw2}
G.~F.~Giudice, R.~Rattazzi and J.~D.~Wells,
Nucl.\ Phys.\ B {\bf 595}, 250 (2001).

\bibitem{peskin} S.~Cullen, M.~Perelstein and M.~E.~Peskin,
Phys.\ Rev.\ D {\bf 62}, 055012 (2000);
E.~Dudas and J.~Mourad,
Nucl.\ Phys.\ B {\bf 575}, 3 (2000);
E.~Accomando, I.~Antoniadis and K.~Benakli,
Nucl.\ Phys.\ B {\bf 579}, 3 (2000);


\bibitem{perry}
R.~C.~Myers and M.~J.~Perry,
Annals Phys.\  {\bf 172}, 304 (1986).

\bibitem{landau}
L. D. Landau and E.M. Lifshitz, Quantum Mechanics, vol. 3 of
{\it Course of Theoretical Physics}.

\bibitem{thooft} G.~'t~Hooft, Phys.\ Lett.\ B {\bf 198}, 61 (1987).

\bibitem{amati} 
D.~Amati, M.~Ciafaloni and G.~Veneziano,
Phys.\ Lett.\ B {\bf 197}, 81 (1987);
Int.\ J.\ Mod.\ Phys.\ A {\bf 3}, 1615 (1988);
Nucl.\ Phys.\ B {\bf 347}, 550 (1990); Phys.\ Lett.\ B {\bf 289}, 87 (1992).
Nucl.\ Phys.\ B {\bf 403}, 707 (1993).

\bibitem{muzinich} I.~J.~Muzinich and M.~Soldate,
Phys.\ Rev.\ D {\bf 37}, 359 (1988).




\bibitem{verlinde} H.~Verlinde and E.~Verlinde,
Nucl.\ Phys.\ B {\bf 371}, 246 (1992).


\bibitem{eik} H. Cheng and T.T. Wu, 
Phys.\ Rev.\ Lett.\  {\bf 22}, 666 (1969);
H.~Abarbanel and C.~Itzykson,
Phys.\ Rev.\ Lett.\  {\bf 23}, 53 (1969);
M.~Levy and J.~Sucher, Phys.\ Rev.\ {\bf 186}, 1656 (1969).  

\bibitem{kab1} D.~Kabat and M.~Ortiz,
Nucl.\ Phys.\ B {\bf 388}, 570 (1992);
D.~Kabat,
Comments Nucl.\ Part.\ Phys.\  {\bf 20}, 325 (1992).


\bibitem{Eardley:2002re}
D.~M.~Eardley and S.~B.~Giddings,
arXiv:gr-qc/0201034;
E.~Kohlprath and G.~Veneziano,
arXiv:gr-qc/0203093.

\bibitem{hawkingellis} S.W. Hawking and G.R.F. Ellis, {\it The large scale structure of space-time}
(Cambridge UNiversity Press, 1973).

\bibitem{d'eathpayne}
P.~D.~D'Eath and P.~N.~Payne,
Phys.\ Rev.\ D {\bf 46}, 658 (1992); {\it ibid.} 675; {\it ibid.} 694.

\bibitem{voloshin}
M.~B.~Voloshin, Phys.\ Lett.\ B {\bf 518}, 137 (2001);
hep-ph/0111099.

\bibitem{banks}
T.~Banks and W.~Fischler,
hep-th/9906038.

\bibitem{blackh} 
S.~B.~Giddings and S.~Thomas,
hep-ph/0106219;
S.~Dimopoulos and G.~Landsberg,
Phys.\ Rev.\ Lett.\  {\bf 87}, 161602 (2001);
S.~B.~Giddings,
hep-ph/0110127.

\bibitem{empa} 
R.~Emparan,
Phys.\ Rev.\ D {\bf 64}, 024025 (2001).

\bibitem{ratta}
R.~Emparan, M.~Masip and R.~Rattazzi, 
Phys.\ Rev.\ D {\bf 65}, 064023 (2002).


\bibitem{us}
G.~F.~Giudice, R.~Rattazzi and J.~D.~Wells,
Nucl.\ Phys.\ B {\bf 630}, 293 (2002)


\bibitem{vlhc}
See
{\tt http://www-ap.fnal.gov/VLHC/}.

\bibitem{Emparan:2000rs}
R.~Emparan, G.~T.~Horowitz and R.~C.~Myers,
Phys.\ Rev.\ Lett.\  {\bf 85}, 499 (2000)

\bibitem{Feng:2001ib}
J.~L.~Feng and A.~D.~Shapere,
Phys.\ Rev.\ Lett.\  {\bf 88}, 021303 (2002); L.~Anchordoqui and H.~Goldberg,
Phys.\ Rev.\ D {\bf 65}, 047502 (2002)

\bibitem{nussi}
S.~Nussinov and R.~Shrock,
Phys.\ Rev.\ D {\bf 64}, 047702 (2001).

\bibitem{Gandhi:1998ri}
R.~Gandhi, C.~Quigg, M.~H.~Reno and I.~Sarcevic,
Phys.\ Rev.\ D {\bf 58}, 093009 (1998)


\bibitem{Ringwald:2001vk}
A.~Ringwald and H.~Tu,
Phys.\ Lett.\ B {\bf 525}, 135 (2002)

\bibitem{Anchordoqui:2001cg}
L.~A.~Anchordoqui, J.~L.~Feng, H.~Goldberg and A.~D.~Shapere,
arXiv:hep-ph/0112247.



\bibitem{Weiler:2000ku}
T.~J.~Weiler,
AIP Conf.\ Proc.\  {\bf 579}, 58 (2001)
[arXiv:hep-ph/0103023].













\bibitem{cteq}
H.~L.~Lai {\it et al.},
Phys.\ Rev.\ D {\bf 51}, 4763 (1995).

\bibitem{grv}
M.~Gluck, E.~Reya and A.~Vogt,
Z.\ Phys.\ C {\bf 67}, 433 (1995).

\bibitem{pythia}
T.~Sjostrand, L.~Lonnblad and S.~Mrenna,
hep-ph/0108264.


\bibitem{benakli}
I.~Antoniadis, K.~Benakli and A.~Laugier,
JHEP {\bf 0105}, 044 (2001).


\bibitem{dimemp} 
S.~Dimopoulos and R.~Emparan,
hep-ph/0108060.

\end{thebibliography}
\end{document}